\documentclass[%
 reprint,
 amsmath,amssymb,
 aps,
 nofootinbib
]{revtex4-2}

\usepackage{graphicx}
\usepackage{dcolumn}
\usepackage{bm}
\usepackage{tabularx}
\usepackage{booktabs}   
\usepackage{multirow}   
\usepackage{array}      
\usepackage{xcolor}
\usepackage{hyperref}
\newcommand{\SC}{\ensuremath{{\rm SC}}}
\newcommand{\NSC}{\ensuremath{{\rm NSC}}}
\newcommand{\AC}{\ensuremath{{\rm AC}}}
\newcommand{\NAC}{\ensuremath{{\rm NAC}}}
\newcommand{\llangle}{\ensuremath{\langle\langle}}
\newcommand{\rrangle}{\ensuremath{\rangle\rangle}}

\usepackage{xcolor}

\begin{document}

\preprint{APS/123-QED}

\title{Probing Nuclear Geometry through Multi-Particle Azimuthal Correlations and Rapidity-Even Dipolar Flow in ${}^{16}$O+${}^{16}$O Collisions}

\author{Kaiser Shafi}
\email{kaisers@iiserbpr.ac.in}
\author{Sandeep Chatterjee}
\email{sandeep@iiserbpr.ac.in}

\affiliation{Department of Physical Sciences, Indian Institute of Science Education and Research Berhampur,\\ Laudigam--760003, Dist.--Ganjam, Odisha, India}
\date{\today}

\begin{abstract}
We study symmetric and asymmetric cumulants as well as rapidity-even dipolar flow in ${}^{16}$O+${}^{16}$O collisions at $\sqrt{s_{NN}} = 200$~GeV to explore $\alpha$-clustering phenomena in light nuclei within the viscous relativistic hydrodynamics framework. Signatures of $\alpha$-clustering manifest in the anisotropic flow coefficients and their correlations---particularly in observables involving elliptic-triangular flow correlations. We show that final-state symmetric and asymmetric cumulants---especially $\mathrm{NSC}(2,3)$ and $\mathrm{NAC}_{2,1}(2,3)$---are sensitive to the initial nuclear geometry. Additionally, we observe a significant difference in rapidity-even dipolar flow, $v_1^{\text{even}}$, between $\alpha$-clustered and Woods--Saxon configurations in high-multiplicity events. These findings underscore the pivotal role of nuclear structure in heavy-ion collision dynamics and provide observables for distinguishing nuclear geometries, particularly in ultra-central collisions.
\end{abstract}

\maketitle

\section{Introduction}
The study of nuclear geometry has long been essential for understanding the structure and dynamics of atomic nuclei. Evidence for nuclear deformation is generally derived from low-energy spectroscopic measurements and models that assess reduced transition probabilities between low-lying rotational states, utilizing beam energies below a few tens of MeV per nucleon~\cite{Bender:2003jk,Heyde:2011pgw,Moller:2015fba}. This focus on nuclear structure has provided insights into the fundamental forces and interactions that govern matter at the subatomic level~\cite{close2015nuclear}, with ongoing challenges in investigating complex nucleon-nucleon interactions at low energies~\cite{Delaroche:2009fa,Wang:2024kdo}. As part of this, various phenomenological models~\cite{Wang:2024kdo,Demyanova:2024ahe,Hamada:2023sjd,Morales-Gallegos:2022dzq,Kekejian:2022ipn} and experimental methodologies~\cite{Yang:2022wbl,Magdy:2024thf,Cline:1986ik} have been developed, exploring nuclei such as oxygen and neon through first-principles approaches like NLEFT simulations~\cite{Meissner:2014lgi,Elhatisari:2017eno}, PGCM~\cite{Frosini:2021fjf,Frosini:2021sxj,Frosini:2021ddm}, and VMC~\cite{Lonardoni:2018nob}.

In high-energy nuclear physics, one key motivation is to probe the bulk properties of QCD matter under extreme conditions, such as those found in the early universe or in astrophysical objects~\cite{Busza:2018rrf}. A significant example is the quark-gluon plasma (QGP), a hot phase of QCD matter that behaves as a near-perfect fluid~\cite{Teaney:2009qa}. The QGP provides a unique environment for studying the collective behavior of strongly interacting matter, and its properties are inferred from the harmonic spectrum of the azimuthal distributions of final-state hadrons. The coefficients \(v_n\) associated with these distributions quantify anisotropic flow and serve as direct signatures of hydrodynamic behavior, offering critical insights into the early dynamics of the collision~\cite{Ollitrault:2023wjk,Teaney:2010vd}. Understanding of QGP has been central to high-energy nuclear physics, with key experimental results from RHIC and the LHC providing valuable insights into its properties under extreme conditions~\cite{BRAHMS:2004adc,PHOBOS:2004zne,STAR:2005gfr,PHENIX:2004vcz,ALICE:2016fzo,CMS:2016fnw}. These findings support the widely accepted model describing the hydrodynamic expansion of QGP with low shear viscosity~\cite{Heinz:2013th,Romatschke:2017ejr,Berges:2020fwq} and a subsequent late-stage hadronic afterburner for interactions among hadrons.

A unique aspect of relativistic heavy-ion collisions is their ability to probe the initial geometric structures of the colliding nuclei. These collisions provide important information about the spatial distribution and clustering of nucleons within light nuclei, which can be studied through observables like collective flow~\cite{STAR:2013ayu,STAR:2014clz,STAR:2013qio,STAR:2015wza,Broniowski:2013dia,Li:2020vrg,Rybczynski:2017nrx,Prasad:2024ahm,Behera:2023nwj,He:2021uko,Bozek:2014cva,Summerfield:2021oex,Ding:2023ibq,Wang:2021ghq,Lu:2025cni,Lim:2018huo,Rybczynski:2019adt,Liu:2023gun,Giacalone:2024luz}, Hanbury Brown-Twiss (HBT) correlations~\cite{STAR:2014shf,Lacey:2014wqa,PHENIX:2014xme}, and fluctuations~\cite{STAR:2014nuj,Luo:2017faz}. These observables shed light on the initial nuclear geometry, offering valuable clues regarding the properties of the initial-state nuclei, and are essential tools for investigating nuclear structure at high energies. In recent decades, relativistic heavy-ion collisions have become a crucial technique for probing the intrinsic properties of nuclei, including their geometric configurations and cluster structures~\cite{STAR:2024wgy,Giacalone:2023cet,Mantysaari:2023qsq,STAR:2021mii,Giacalone:2020ymy,Filip:2009zz,Shou:2014eya,Giacalone:2019pca,Zhang:2021kxj,Li:2019kkh,Liu:2022kvz}.

The dynamics of nuclear collisions are now widely understood through the framework of relativistic hydrodynamics, where the collective behavior of the system is driven by the initial collision geometry~\cite{Teaney:2010vd,Gardim:2011xv,Niemi:2012aj,Teaney:2012ke,Qiu:2011iv,Gardim:2014tya,Betz:2016ayq,Rao:2019vgy,Hippert:2020kde}. Precision observables for probing the hydrodynamic evolution of these collisions, such as multi-particle correlations~\cite{Bilandzic:2010jr,Luzum:2013yya,Zhou:2016eiz}, soft-hard multi-particle azimuthal correlations~\cite{Betz:2016ayq,Prado:2016szr,ALICE:2018gif}, have become increasingly important in recent years, driving the field into a new era of precision physics~\cite{Noronha-Hostler:2015dbi,Niemi:2015voa,ALICE:2015juo,Bernhard:2019bmu,Shen:2020mgh}.

\subsection{Symmetric and Asymmetric Cumulants of Flow Amplitudes}
Anisotropies observed in the momentum distributions of produced particles carry information about the QGP properties~\cite{Heinz:2013th,Braun-Munzinger:2015hba,Busza:2018rrf}. After hadronization, particles are emitted anisotropically in the transverse plane, quantified using Fourier series of flow amplitudes $v_n$ and symmetry planes $\psi_n$~\cite{Voloshin:1994mz},
\begin{equation}
     \frac{dN}{d\phi} = \frac{1}{2\pi} \left[ 1 + 2 \sum_{n=1}^{\infty} v_n \cos\left[n\left(\phi-\psi_n\right)\right] \right]
\end{equation}
\par
Collective flow provides insight into the QGP medium~\cite{Ollitrault:1992bk, Heinz:2013th,Bass:1998vz}. The impact parameter driven spatial geometry of the fireball and nucleonic fluctuations lead to various anisotropic flow harmonics. The second-order elliptic flow ($v_2$) originates from the geometry of the overlap region~\cite{Voloshin:2008dg, Voloshin:1999gs, Ivanov:2014zqa, STAR:2004jwm, STAR:2001ksn, ALICE:2010suc, Sorge:1998mk}, while the third-order triangular flow ($v_3$) arises from event-by-event nucleon and sub-nucleon fluctuations~\cite{STAR:2013qio, Solanki:2012ne, ALICE:2016cti, Heinz:2013bua, Alver:2010gr}. These flow harmonics are sensitive to the equation of state (EoS) and transport properties of the fireball, such as viscosity~\cite{Parkkila:2021tqq, Retinskaya:2014zea, Shen:2012vn, Teaney:2003kp, Shen:2011zc}.
\par
Recent studies have demonstrated that higher-order observables, such as flow cumulants, provide valuable insights into the properties of the Quark-Gluon Plasma (QGP) by probing genuine correlations between different moments of flow harmonics while remaining robust against non-flow effects, as evidenced by simulations using the HIJING Monte Carlo generator~\cite{PhysRevC.105.024912}. These developments stem from advancements in multi-particle correlation analysis, which have enabled the assessment of correlators involving different harmonics~\cite{Bilandzic:2010jr,Bilandzic:2013kga,Mordasini:2019hut,ALICE:2017kwu}. Symmetric and asymmetric cumulants, in particular, have emerged as a powerful tools for quantifying correlations between event-by-event fluctuations in flow harmonics \( v_m \) and \( v_n \), allowing for the differentiation of true collective flow from non-flow contributions. Measurements by the STAR and the ALICE collaborations have revealed significant correlations among \( v_2 \), \( v_3 \), and \( v_4 \), and have shown sensitivity to the shear viscosity of the medium~\cite{STAR:2013qio, Solanki:2012ne, ALICE:2016cti, Heinz:2013bua, Alver:2010gr}.

The symmetric cumulant, $\SC(m,n)$, is defined as~\cite{Bilandzic:2013kga}
\begin{eqnarray}
\SC(m,n) &\equiv& \langle v_m^2 v_n^2 \rangle _c \nonumber\\
        &=& \left<v_{m}^2v_{n}^2\right>-\left<v_{m}^2\right>\left<v_{n}^2\right>\nonumber\\
        &=& \left<\left<\cos(m\varphi_1\!+\!n\varphi_2\!-\!m\varphi_3-\!n\varphi_4)\right>\right>\nonumber\\
        && - \left<\left<\cos[m(\varphi_1\!-\!\varphi_2)]\right>\right>\left<\left<\cos[n(\varphi_1\!-\!\varphi_2)]\right>\right>\nonumber\\
\label{scmn_1}
\end{eqnarray}
The subscript \( c \) denotes that this quantity is a cumulant. A positive or negative value for \( \SC(m,n) \) indicates whether there is a correlation or anti-correlation between \( v_m^2 \) and \( v_n^2 \). Specifically, when \( v_m^2 \) exceeds its average value in an event, the chances that \( v_n^2 \) similarly exceeds its average value rise (or fall) accordingly. This focus on correlations across different flow harmonics enables effective comparisons between experimental results and theoretical predictions.

To mitigate the influence of the magnitudes of \( v_m \) and \( v_n \) on the symmetric cumulant value, \( \SC(m,n) \) is normalized using their average values, \( \langle v_m^2 \rangle \) and \( \langle v_n^2 \rangle \). This normalization leads to what is termed the normalized symmetric cumulant, \( \NSC(m,n) \), facilitating straightforward comparisons of experimental data with theoretical models as well as between initial and final state fluctuations~\cite{Mordasini:2019hut}.
The normalized symmetric cumulant, $\NSC(m,n)$, is given by~\cite{Taghavi:2020gcy}:
\begin{equation}
\NSC(m,n) = \frac{\SC(m,n)}{\langle v_{m}^{2} \rangle \langle v_{n}^{2} \rangle} 
\label{scmn_norm1}
\end{equation}

Asymmetric cumulants capture genuine correlations between different moments of flow harmonics. The asymmetric cumulant, $\AC_{2,1}(m,n)$, is defined as~\cite{PhysRevC.105.024912}:
\begin{equation}
    \begin{split}
    \label{eq_ac21}
        \AC_{2,1}(m,n) &  \equiv \langle (v_m^2)^2 v_n^2 \rangle _c \\
        & = \langle v_m^4 v_n^2 \rangle - \langle v_m^4 \rangle \langle v_n^2 \rangle 
        - 2 \langle v_m^2 v_n^2 \rangle \langle v_m^2 \rangle 
        + 2 \langle v_m^2 \rangle^2 \langle v_n^2 \rangle \\
        & = \llangle {\rm e}^{i(m\varphi_1 + m\varphi_2 + n\varphi_3 - m\varphi_4 - m\varphi_5 - n\varphi_6)} \rrangle\\
        & \quad - \llangle {\rm e}^{i(m\varphi_1 + m\varphi_2 - m\varphi_3 - m\varphi_4)} \rrangle \llangle {\rm e}^{i(n\varphi_1 - n\varphi_2)} \rrangle \\
        & \quad - 2  \llangle {\rm e}^{i(m\varphi_1 + n\varphi_2 - m\varphi_3 - n\varphi_4)} \rrangle \llangle {\rm e}^{i(m\varphi_1 - m\varphi_2)} \rrangle \\
        & \quad + 2  \llangle {\rm e}^{i(m\varphi_1 - m\varphi_2)} \rrangle^2 \llangle {\rm e}^{i(n\varphi_1 - n\varphi_2)} \rrangle
    \end{split}
\end{equation}
In this equation, the subscripts $(2,1)$ signify the powers of $v_m^2$ and $v_n^2$ in the cumulant \(\langle (v_m^2)^2 v_n^2 \rangle _c\).
These equations denote genuine multivariate cumulants. Additionally, the asymmetric cumulant $\AC_{1,1}(m,n)$ equates to the symmetric cumulant $\SC(m,n)$, highlighting a broader application of the asymmetric cumulants. The normalization of these cumulants provides significant advantages, such as facilitating meaningful comparisons of results and allowing for the identification of initial state effects and changes resulting from hydrodynamic evolution. Since the scales of AC predictions differ between initial and final states, normalization aids in reconciling these disparities. Moreover, as flow amplitudes vary with transverse momentum ($p_T$), normalization negates the $p_T$ dependence present in linear combinations of these amplitudes, like symmetric and aysmmetric cumulants, thus enabling comparisons across different models and data sets with a range of $p_T$ values.\begin{eqnarray}
\label{NACDefinition}
\NAC_{2,1}(m,n) = \frac{{\AC}_{2,1}(m,n)}{\langle v_m^2 \rangle^2 \langle v_n^2 \rangle}
\end{eqnarray}
\par
We computed both normalized symmetric and asymmetric cumulants for ${}^{16}$O + ${}^{16}$O collisions at $\sqrt{s_{NN}} = 200$ GeV. Previous studies have explored these cumulants in heavy-ion collisions, such as Au+Au and Pb+Pb, employing various hydrodynamic and hadronic transport models~\cite{Zhu:2016puf, Schenke:2019ruo, Mordasini:2019hut, Taghavi:2020gcy, Hirvonen:2022xfv, Magdy:2022ize, Magdy:2024ooh, Gardim:2016nrr,Nasim:2016rfv}. Additionally, experiments conducted by the STAR experiment at RHIC~\cite{2018459} and the ALICE experiment at LHC~\cite{ALICE:2016kpq,ALICE:2023lwx} have measured these cumulants. The ALICE collaboration has performed a systematic study of $\SC(m,n)$ and $\NSC(m,n)$~\cite{ALICE:2017kwu}. Their findings show a positive correlation between \( v_2 \) and \( v_4 \), while indicating a negative correlation between \( v_2 \) and \( v_3 \). In addition to symmetric and asymmetric cumulants, various other correlators have also been studied~\cite{ATLAS:2018ngv,Ortiz:2019osu,CMS:2019lin,ALICE:2021adw}.

We do not provide predictions for mixed harmonic cumulants (MHC) in this work~\cite{Moravcova:2020wnf}. As shown in Ref.~\cite{PhysRevC.105.024912}, the MHC formalism does not yield valid cumulants of the flow amplitudes, and therefore its use is not appropriate for the present analysis.

\subsection{\( v_1^{\rm even} \)}
Directed flow can be characterized as either rapidity-even or rapidity-odd~\cite{Gardim:2011qn}. The odd component, \( v_1^{\rm odd}(y) \), originating from the geometry of the overlap region of the collision and associated with the reaction plane, has been extensively studied~\cite{STAR:2003xyj,NA49:2003njx,STAR:2005btp,PHOBOS:2005ylx,STAR:2008jgm,ALICE:2013xri,Parida:2025lhn,Parida:2025ddt,Parida:2023tdx,Parida:2023rux,Parida:2023ldu,Parida:2022zse,Parida:2022ppj}. In symmetric collisions, the even component, \( v_1^{\rm even}(y) \), arises from fluctuations in the initial nucleon positions and varies weakly with rapidity~\cite{Bozek:2010vz,Bozek:2012hy}. These fluctuations break the symmetry of the initial density profile and generate a preferred direction, where the gradient is steepest. This anisotropy in the initial state is quantified by the dipole asymmetry:
\begin{equation}
\label{eq_e1}
\varepsilon_1 e^{i\Phi_1} = - \frac{\langle r^3 e^{i\phi} \rangle}{\langle r^3 \rangle},
\end{equation}
where the averages are taken over the initial transverse entropy density profile, with \((r, \phi)\) as polar coordinates centered such that \( \langle r e^{i\phi} \rangle = 0 \). 
Experimentally, $v_1$ is extracted from a two-particle correlation, which scales like $(v_1)^2$, so the measured $v_1$ should scale with the root-mean-square (rms) dipole asymmetry, $\varepsilon_1\{2\} \equiv \sqrt{\langle \varepsilon_1^2 \rangle}$~\cite{Miller:2003kd}. Due to fluctuations, \( \varepsilon_1 \) is non-zero even at midrapidity. Hydrodynamic expansion of the dipole asymmetry, \(\varepsilon_1\), generates directed flow anisotropy in the momentum distribution of the emitted particles. High-\(p_T\) particles are preferentially emitted in the direction of the steepest gradient (\( \Psi_1 \)), where the fluid velocity is the largest~\cite{Borghini:2005kd}. This anisotropy is characterized by the directed flow with respect to \( \Psi_1 \):
\begin{equation}
v_1 = \langle \cos(\phi - \Psi_1) \rangle,
\end{equation}
where \( \phi \) is the particle azimuthal angle, and the average is taken over particles in each event. Because total transverse momentum must approximately vanish, high-\(p_T\) particles with positive \(v_1\) imply a corresponding negative \(v_1\) for low-\(p_T\) particles, resulting in a characteristic \(p_T\)-dependence of \(v_1\)~\cite{Teaney:2010vd}.

Rapidity even directed flow is typically analyzed via the event-plane method, analogous to that used for elliptic flow~\cite{Danielewicz:1985hn,Poskanzer:1998yz}. In each event, the event plane angle \( \Psi_1 \) is defined as~\cite{Luzum:2010fb}:
\[
Q e^{i\Psi_1} = \sum_{j} w_j e^{i\phi_j},
\]
where the sum runs over charged particles in an event. The weight \( w_j \) for each particle is selected such that \( \langle w p_T \rangle = 0 \) to prevent bias from momentum conservation, as these correlations are proportional to \( p_T \)~\cite{Jia:2012hx,Borghini:2000sa}. For the even component, the weight is chosen to depend on $p_T$, as follows~\cite{Luzum:2010fb}:
\[
w(p_T) = p_T - \frac{\langle\langle p_T^2 \rangle\rangle}{\langle\langle p_T \rangle\rangle},
\]
where the inner average is over particles in an event, and the outer over all events. This choice satisfies \( \langle w p_T \rangle = 0 \), suppressing momentum conservation bias. 
The final $Q$-weighted directed flow coefficient is given by:
\[
v_1^{\rm even} = \frac{\langle Q \cos(\phi - \Psi_1) \rangle}{\sqrt{\langle Q^2 \rangle}}.
\]

The Fourier harmonics \( v_n \) probe increasingly smaller spatial scales with larger \( n \), making higher harmonics more sensitive to viscosity. In contrast, \( v_1 \) is relatively insensitive to viscous effects~\cite{Retinskaya:2012ky,Luzum:2008cw,Alver:2010dn,Schenke:2011bn}, and its proportionality to \( \varepsilon_1 \) makes it a direct probe of the system’s early-time dipole asymmetry~\cite{Magdy:2023fsp}. Measurements of $v_1^{\rm even}$ have been reported in Au+Au collisions by the STAR Collaboration at RHIC~\cite{STAR:2018gji}, and in Pb+Pb collisions by the ALICE and ATLAS Collaborations at the LHC~\cite{ALICE:2013xri,Jia:2012hx}.

\section{Framework}
In this work, we utilize a multi-component framework to simulate different stages of heavy ion collisions. The initial energy density profiles are generated using the TRENTo model. We compute the nuclear distribution using the three-parameter Fermi (3pF) distribution and tetrahedral configurations of $\alpha$ clusters (with different compactness of the $\alpha$ clusters in the nuclei) with the root-mean-square radius for $^{16}$O kept constant. The 3pF nuclear distribution is given by:
\[
\rho = \rho_0 \left(1 + \omega r^2 / R^2 \right) \left[1 + \exp\left((r - R) / a\right)\right]^{-1},
\]

In the tetrahedral configurations, the shape of the oxygen nucleus is described by a tetrahedron with side length $l$, and the centers of the four $\alpha$ clusters are positioned at the vertices of the tetrahedron. The spatial coordinates of the nucleons in each $\alpha$ cluster are sampled from a 3D Gaussian distribution with a parameter $r_\alpha$, which represents the root-mean-square radius of each cluster. The parameter $r_\alpha$ reflects the compactness of the $\alpha$ cluster in the nucleus: a smaller value of $r_\alpha$ corresponds to a denser cluster.

\begin{figure}[]
\includegraphics[width=0.23\textwidth]{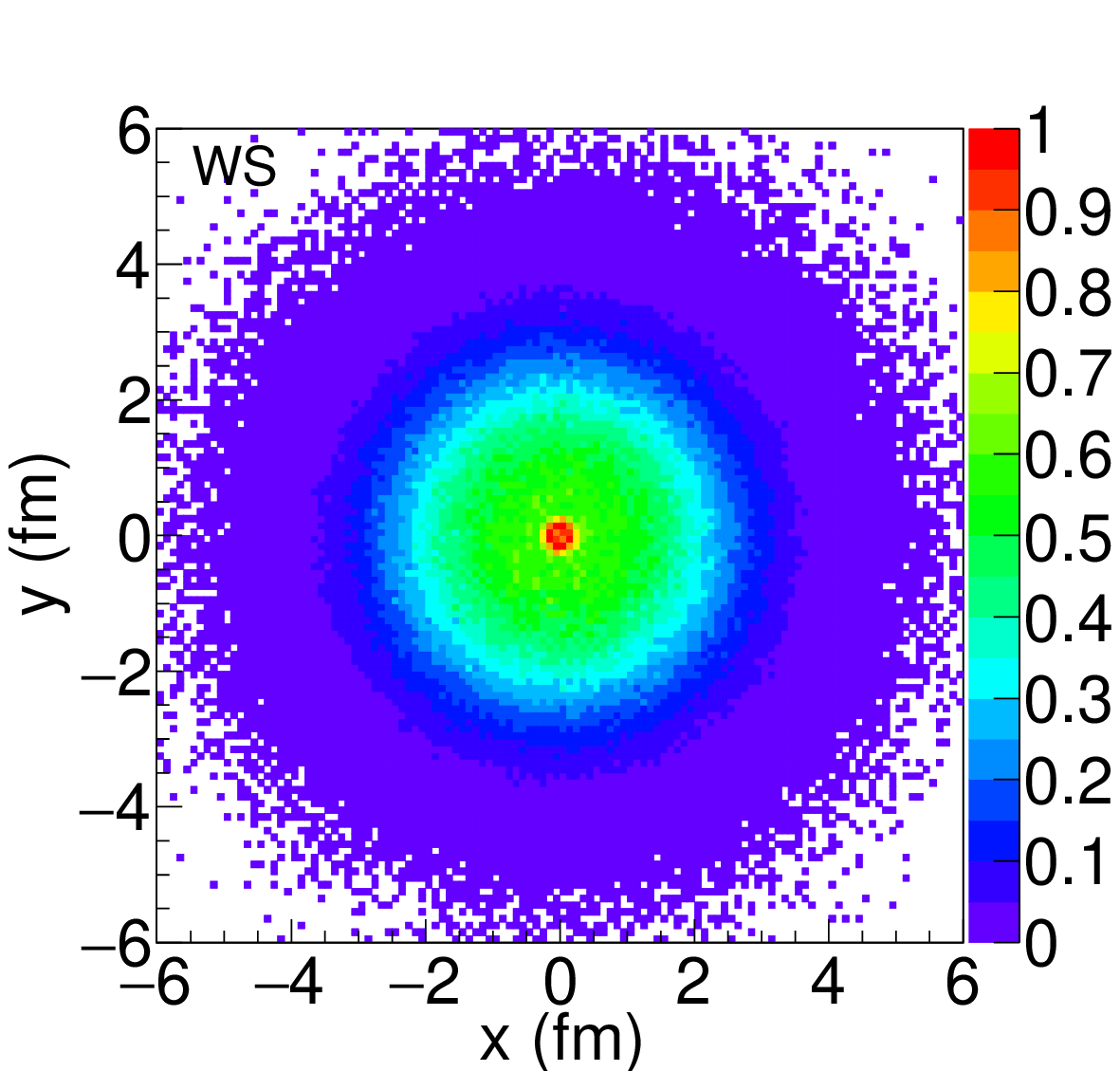}
\includegraphics[width=0.23\textwidth]{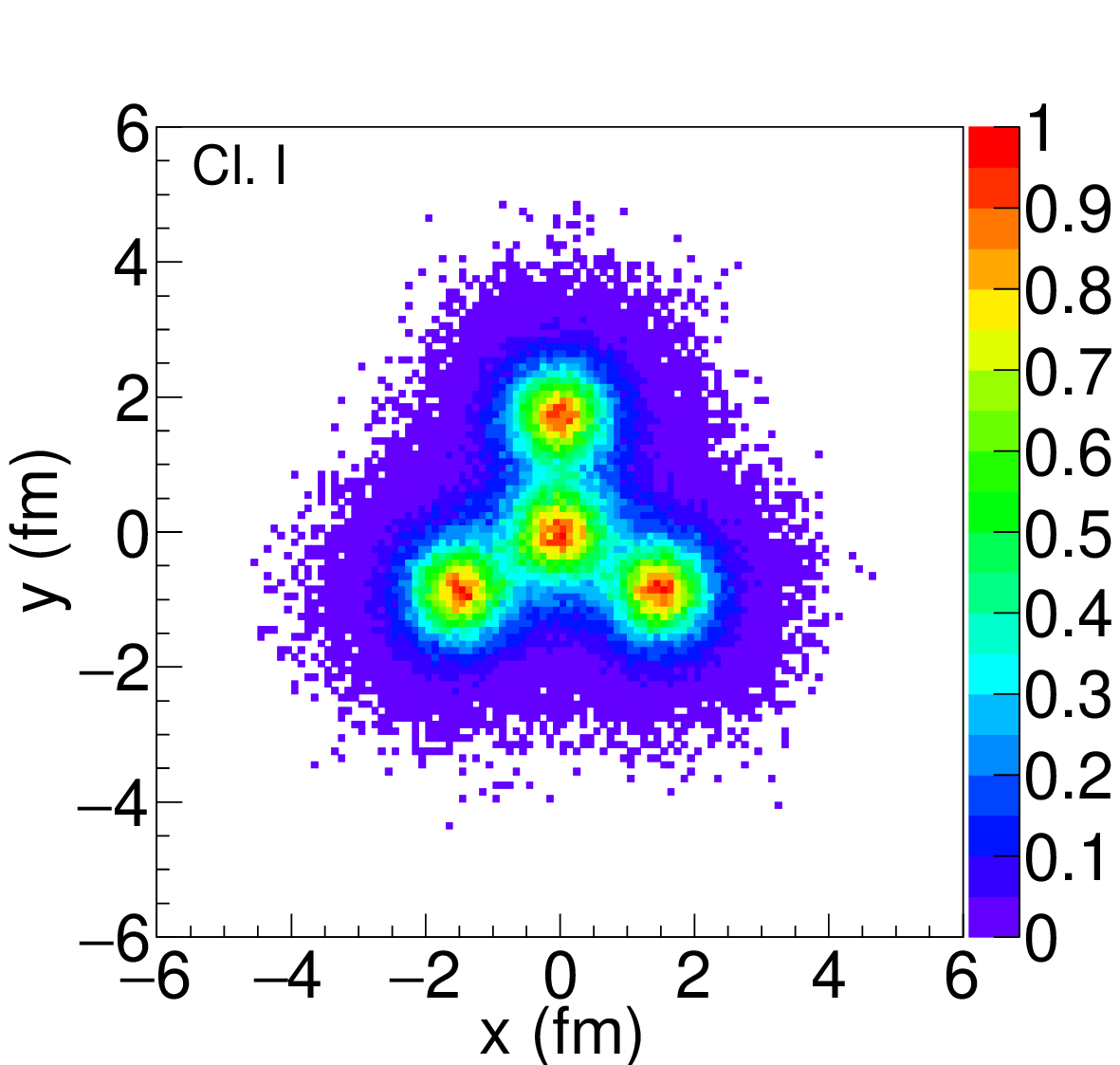}\\
\includegraphics[width=0.23\textwidth]{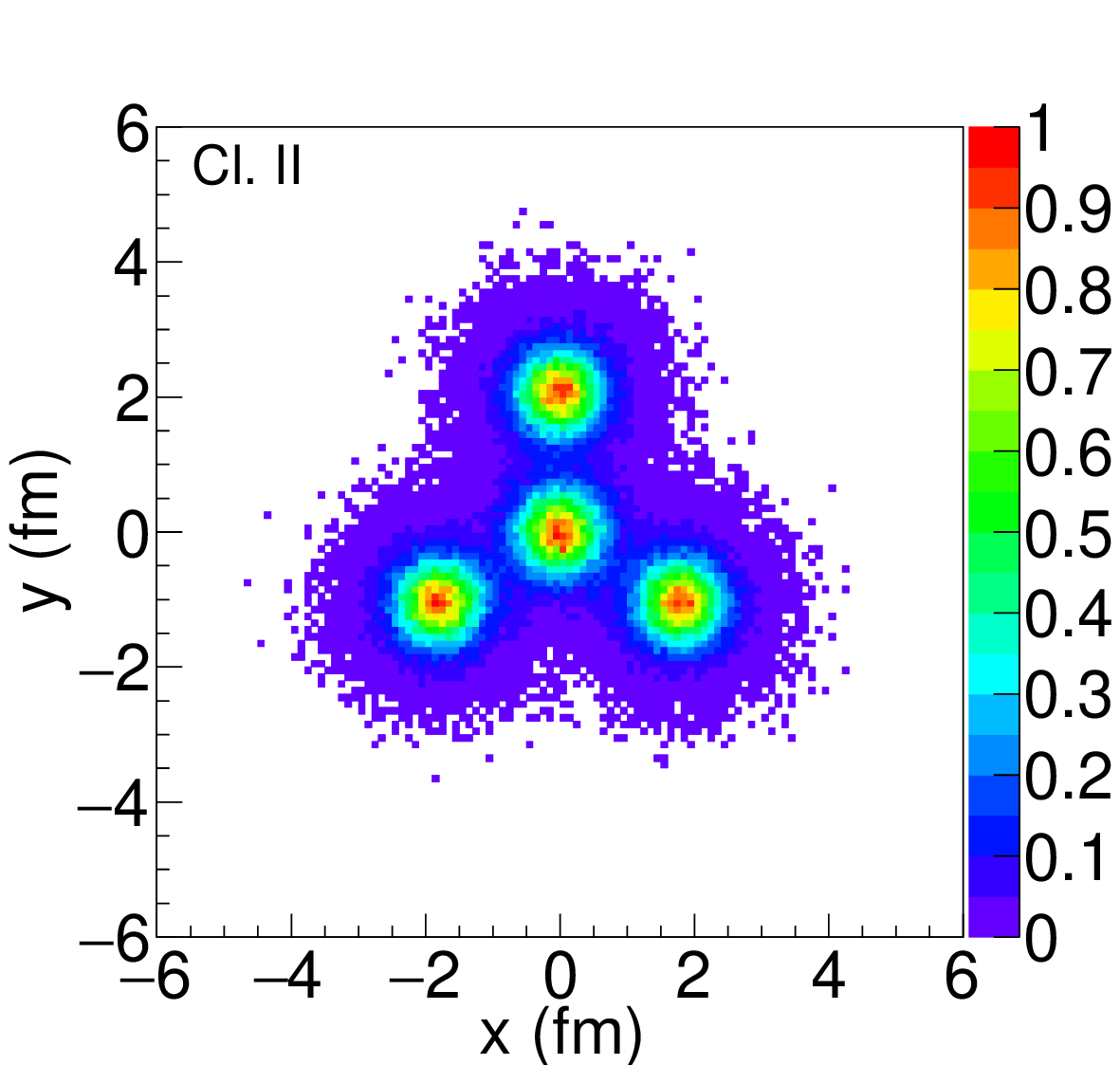}
\includegraphics[width=0.23\textwidth]{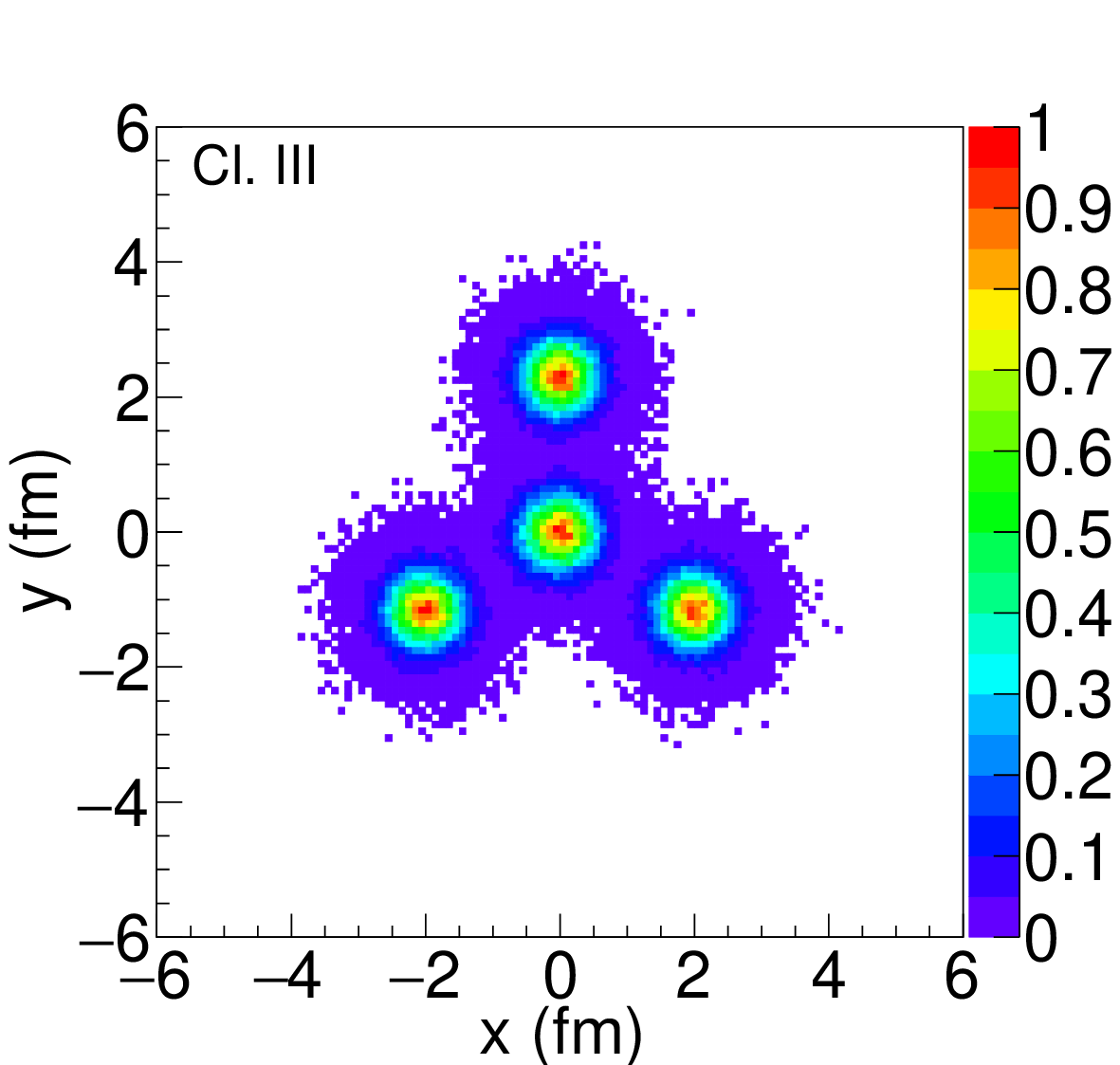}
\caption{\label{plot:nuc_dist} Nuclear distributions of ${}^{16}$O using Woods--Saxon and tetrahedral $\alpha$-cluster configurations (see Table~\ref{tab:ic_par}).}
\end{figure}

In this study, the nuclear density of $^{16}$O is constructed using three different values of $r_\alpha$, taken from the reference~\cite{YuanyuanWang:2024sgp}, to comprehensively understand the effect of $\alpha$ cluster configurations on the final observables. We impose that the root-mean-square radius of $^{16}$O should be the same for the different densities, i.e., 
\[
\sqrt{\langle r^2 \rangle} \equiv \sqrt{\frac{3l^2}{8} + r_{\alpha}^2} = 2.73 \, \text{fm},
\]
which is taken from nuclear structure experiments~\cite{DeVries:1987atn}. The differences between charge density and nuclear mass density are ignored. In the case of a more compact $\alpha$ cluster, a larger value of $l$ is required. At small values of \( r_\alpha / l \), distinct configurations of hotspots---three at the vertices of a triangle, with a fourth at the center---emerge. The nuclear distribution parameters are listed in Table~\ref{tab:ic_par}, and the two-dimensional density profiles, projected into the transverse plane of the collision, are shown in Fig.~\ref{plot:nuc_dist}.

The parameters---reduced thickness parameter ($p$), gamma distribution shape parameter ($k$), nucleon width ($w_{\rm c}$), and minimum nucleon-nucleon distance ($d_{\rm min}$)
, in TRENTo model for the simulation of ${}^{16}$O+${}^{16}$O collisions at $\sqrt{s_{NN}} = 200$ GeV are given in Table~\ref{table:trento_param}.

\begin{table}
    \caption{The parameters~\cite{DeVries:1987atn} for the nuclear distributions of ${}^{16}$O with Woods--Saxon and tetrahedral configurations of $\alpha$ clusters.}
    \label{tab:ic_par}
    \centering
    \resizebox{0.45\textwidth}{!}{  
    \small 
    \begin{tabular}{p{1cm} p{2.5cm} p{1.3cm} p{1.3cm} p{1.3cm}}
    \hline
    & Distribution & $R$ & $a$ & $\omega$ \\  
    \hline
    WS    & Woods--Saxon       &   2.608 fm   &  0.513 fm   &  -0.051   \\
    \hline
    \hline
    &  & $l$ & $r_\alpha$ & $r_\alpha / l$ \\  
    \hline
    Cl. I  & $\alpha$ cluster      & 3.0  & 2.0  & 0.67  \\  
    Cl. II   & $\alpha$ cluster      & 3.6  & 1.6  & 0.44  \\ 
    Cl. III   & $\alpha$ cluster      & 4.0  & 1.2  & 0.30  \\  
    \hline
    \end{tabular}
    }
\end{table}

\begin{table}[t!]
\caption{TRENTo model parameters used for ${}^{16}$O+${}^{16}$O collisions, taken from Refs.~\cite{Summerfield:2021oex,Moreland:2018gsh}.}
\centering
\begin{tabular}{|p{0.15\textwidth}|p{0.07\textwidth}|p{0.07\textwidth}|p{0.07\textwidth}|p{0.07\textwidth}|p{0.07\textwidth}|}
\hline
\multicolumn{1}{|c|}{Parameters} & norm & $p$ & $k$ & $w_c$ & $d_{min}$ \\
\hline
\multicolumn{1}{|c|}{Value} & 1.0 & 0.0 & 1.0 & 0.51 fm & 0.4 fm \\
\hline
\end{tabular}
\label{table:trento_param}
\end{table}


The evolution of the energy-momentum tensor starts at an initial time of \( \tau_0 = 0.4 \, \text{fm} \). The hydrodynamic equations are solved using the MUSIC code~\cite{Schenke:2010nt, Schenke:2010rr, Paquet:2015lta, Huovinen:2012is}, which employs the Kurganov-Tadmor algorithm. We initialize the evolution using the entropy density profile from TRENTo, set the $s_{\text{factor}}$ in MUSIC to 89.0, and adjust the TRENTo parameters to the values given in Table~\ref{table:trento_param} to reproduce the Au+Au multiplicity at $\sqrt{s_{NN}} = 200$ GeV. We set a constant effective shear viscosity to entropy density ratio \( \eta/s = 0.08 \), calibrated to match the measured anisotropic flow coefficients of charged hadrons in Au+Au collisions at the same collision energy. A detailed description of the chosen parameters' values is available in Refs.~\cite{Moreland:2018gsh, Summerfield:2021oex}. The specific bulk viscosity is treated as a temperature-dependent quantity and is parametrized as:

\begin{equation}
\frac{\zeta}{s} =
\begin{cases} 
\lambda_1 e^{-(x-1)/\sigma_1} + \lambda_2 e^{-(x-1)/\sigma_2} + 0.001 & \text{for } T > 1.05 T_C, \\
\lambda_3 e^{(x-1)/\sigma_3} + \lambda_4 e^{(x-1)/\sigma_4} + 0.03 & \text{for } T < 0.995 T_C,
\end{cases}
\end{equation}

where \( x = T / T_C \). The parameters \( \lambda_1 = \lambda_3 = 0.9 \), \( \lambda_2 = 0.25 \), \( \lambda_4 = 0.22 \), \( \sigma_1 = 0.025 \), \( \sigma_2 = 0.13 \), \( \sigma_3 = 0.0025 \), and \( \sigma_4 = 0.022 \) are fitted as described in~\cite{Denicol:2009pe}.

For the equation of state (EoS), we employ the NEoS-B model~\cite{Monnai:2019hkn}, which is based on continuum-extrapolated lattice calculations at zero net baryon chemical potential~\cite{HotQCD:2014kol, HotQCD:2012fhj, Ding:2015fca} and smoothly matched to a hadron resonance gas EoS in the temperature range between 110 and 130 MeV~\cite{Moreland:2015dvc}.

We defined centrality classes using the initial entropy, as is common in heavy-ion studies. A more accurate procedure based on final multiplicity would require extensive minimum-bias hydrodynamic simulations, which are currently beyond our computational resources. Nevertheless, our key results are largely insensitive to centrality and remain flat across the most central region. For each nuclear configuration of ${}^{16}$O, we performed hydrodynamic evolution on 5000 initial conditions for each of the centrality classes: 0-0.1\%, 0-1\%, 0-5\%, 20-30\%, and 40-50\%, using two distinct parameter sets (Table~\ref{table:parsets}) for the simulations. A freeze-out hypersurface was constructed at a constant energy density of \( \epsilon_{f} = 0.26 \, \text{GeV}/\text{fm}^3 \), corresponding to a local temperature of approximately 151~MeV. We calculate the invariant single-particle momentum distribution using the Cooper-Frye prescription~\cite{Cooper:1974mv} applied to this hypersurface which is continuous and does not require sampling, unless modeling subsequent late-stage hadronic interactions the effects of which we also discuss, and subsequently include the effects of hadronic resonance decays when constructing the final momentum distribution. As a result, there are no fluctuations due to finite multiplicity, and no statistical error arises from hadronization. This approach allows for achieving reasonable statistical accuracy with a much smaller number of events compared to typical experiments~\cite{Gardim:2011qn}. From this distribution, we compute observables such as flow coefficients and mean transverse momentum. Finally, we compute both symmetric and asymmetric cumulants from the flow magnitudes.

\begin{table}[t!]
\caption{Values of the transport coefficients used in the simulations for the different parameter sets.}
\label{table:parsets}
\centering
\begin{tabular}{|p{0.15\textwidth}|p{0.15\textwidth}|p{0.15\textwidth}|}
\cline{1-3}
  Parameter Sets & $\eta/s$ & $\zeta/s$ \\
  \hline
 \multicolumn{1}{|c|}{I} & 0.08 & 0\\
 \hline
 \multicolumn{1}{|c|}{II} & 0.12 & 0 \\
 \hline
 \multicolumn{1}{|c|}{III} & 0.08 & $\zeta/s(T)$ \\
 \hline
\end{tabular}
\end{table}

The normalized symmetric cumulants, \( \NSC(m,n) \), normalized asymmetric cumulants, \( \NAC_{2,1}(m,n) \), and rapidity-even dipolar flow, $v_1^{even}$, are computed using pseudorapidity (\( \eta \)) ranges of [-0.5, 0.5] and transverse momentum (\( p_T \)) ranges of [0.01, 3.0] GeV. Statistical uncertainties for ensemble-level observables were estimated using the bootstrap method with resampling, which provides a robust estimate of sampling variance without relying on analytical error propagation. The systematic uncertainty bands shown in this work reflect variations in the shear and bulk viscosities only; they therefore provide an indication of the sensitivity of the results to the hydrodynamic stage but do not constitute a full quantitative estimate of the total model uncertainty. Variations in other parameters—such as Trento initial-state parameters or the absence of pre-hydrodynamic evolution —are not included and would require a comprehensive Bayesian analysis to assess reliably.

\section{Results}
\subsection{Mean transverse momentum}
\begin{figure}[t!]
\includegraphics[width=0.45\textwidth]{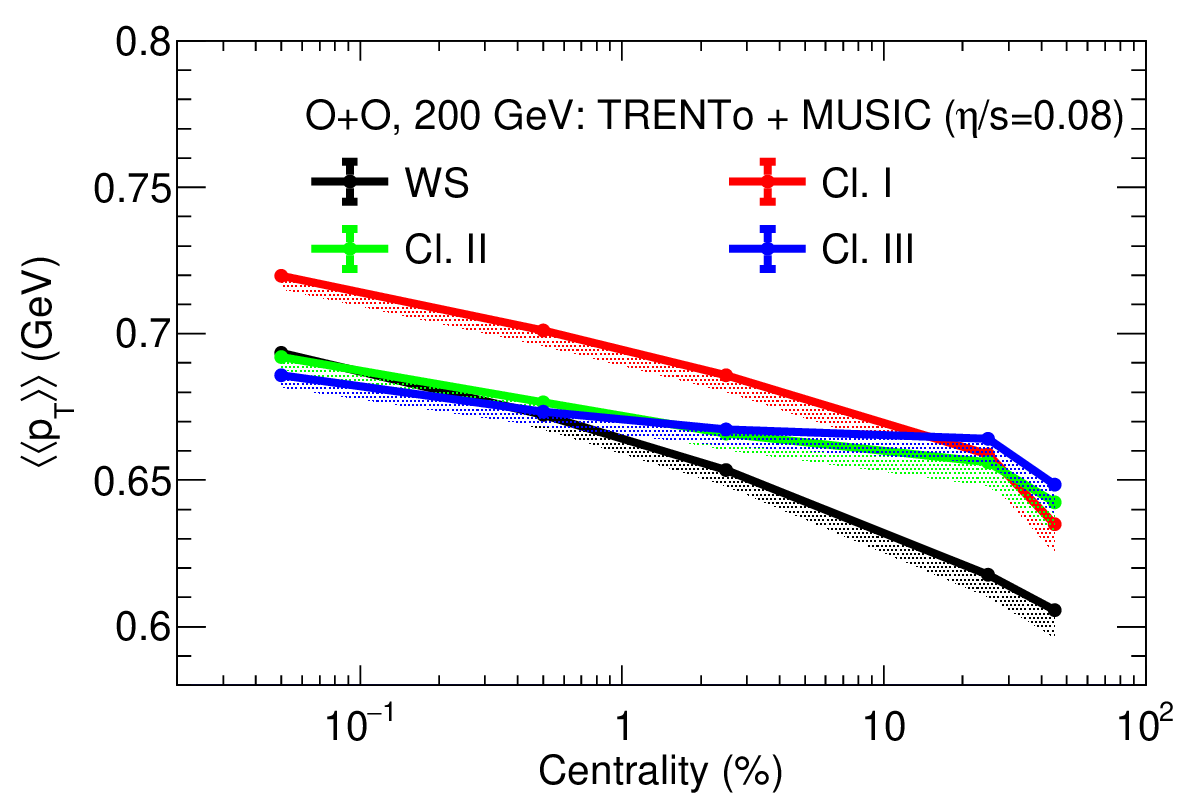}
\caption{\label{plot:mean_pt} $\langle\langle p_T \rangle\rangle$ in ${}^{16}$O+${}^{16}$O collisions at $\sqrt{s_{NN}}$ = 200 GeV for the TRENTo+MUSIC framework. Band represents the systematics due to $\eta/s$.}
\end{figure}

Figure~\ref{plot:mean_pt} shows how the average transverse momentum, $\langle\langle p_{\rm T}\rangle\rangle$, varies with centrality in ${}^{16}$O+${}^{16}$O collisions. In relativistic heavy-ion collisions, $\langle\langle p_{\rm T}\rangle\rangle$ is sensitive to the initial energy-to-entropy (E/S) ratio, $\langle p_{\rm T} \rangle \propto E/S$~\cite{Teaney:2010vd,Giacalone:2020dln}, and typically decreases from central to peripheral collisions due to a reduced hydrodynamic response, as evident in Fig.~\ref{plot:mean_pt}.

The flow development in these collisions is governed primarily by two factors: the spatial homogeneity of the nuclear distribution and the compactness of the $\alpha$ clusters. Since the root-mean-square (RMS) radius of ${}^{16}$O is fixed to its experimental value, the increase in cluster compactness from Cl. I to Cl. III necessarily leads to a corresponding increase in inter-cluster separation.
In ultra-central collisions, the geometry of the overlap region reflects the overall shape of the nucleus, whereas in peripheral collisions, it is dominated by the geometry of the individual clusters.

As a result, in ultra-central collisions, increasing compactness and inter-cluster distance leads to reduced homogeneity and more localized sources of flow (hotspots), causing $\langle\langle p_{\rm T} \rangle\rangle$ to decrease. In contrast, in peripheral collisions, the inter-cluster distance becomes less relevant, and greater compactness enhances $\langle\langle p_{\rm T} \rangle\rangle$ due to larger E/S ratio of individual clusters.

\subsection{Symmetric and Asymmetric Cumulants}
In this section, we present our findings on the normalized symmetric $\NSC(m,n)$ and asymmetric $\NAC_{2,1}(m,n)$ cumulants obtained from our simulations. To evaluate their sensitivity to different nuclear configurations, we compared these observables for the nuclear configurations given in Table~\ref{tab:ic_par}, utilizing the two sets of ensembles detailed in Table~\ref{table:parsets} for TRENTo and TRENTo + MUSIC simulations. Negative values of $\NSC(2, 3)$ throughout the centrality range reveal the anti-correlation between $v_2$ and $v_3$. In contrast, $\NSC(2, 4)$ is positive for all centralities, indicating positive correlations between $v_2$ and $v_4$. Symmetric cumulants do not have contributions from non-flow effects, where non-flow refers to azimuthal correlations not related to the reaction plane orientation, like those from resonances, jets, quantum statistics, etc. This is verified by computing these observables for the HIJING model, which includes only non-flow physics, for which these are consistently zero~\cite{PhysRevC.105.024912}. We found a substantial sensitivity in symmetric and asymmetric cumulants to nuclear geometry which arises primarily from the characteristics of anisotropic flow, as these cumulants incorporate higher powers of flow coefficients. As discussed in the introduction, normalization serves to mitigate the influence of the magnitudes of \(v_m\) and \(v_n\) on the cumulants. Consequently, it reduces the incidental sensitivity of \(\SC(m,n)\) and \(\AC_{2,1}(m,n)\) to shear and bulk viscosities and to resonance decay and late-stage hadronic interactions. Therefore, we show only the results for normalized cumulants, i.e, \(\NSC(m,n)\) and \(\NAC_{2,1}(m,n)\).
\par
\begin{figure}[t!]
\includegraphics[width=0.45\textwidth]{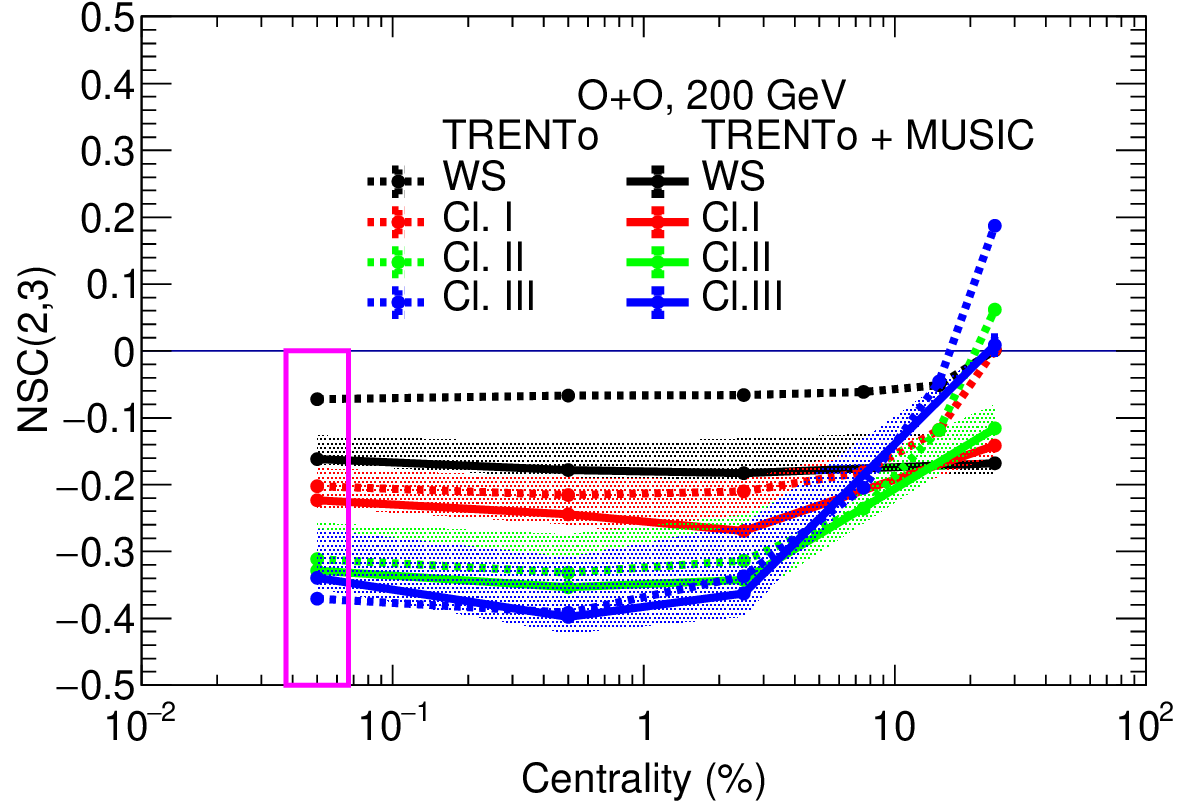}
\caption{\label{plot:NSC_23} Normalized symmetric cumulant vs centrality in ${}^{16}$O+${}^{16}$O collisions at $\sqrt{s_{NN}}$ = 200 GeV. The shaded band represents the systematic uncertainties associated with $\eta/s$ and $\zeta/s$, with the dominant contribution arising from $\zeta/s$ and the influence of $\eta/s$ being comparatively negligible. Dashed lines represent results from TRENTo, while solid lines correspond to the TRENTo+MUSIC framework. For TRENTo+MUSIC, $\NSC(2, 3)$ is consistently negative.}
\end{figure}

Our results indicate that both $\NSC(2,3)$ and $\NSC(2,4)$ are significantly different for the different nuclear configurations. We showed in Ref.~\cite{Shafi:2024tbr} that $\NSC(2,3)$ is particularly insensitive to the hydrodynamic transport coefficients (simulated by MUSIC) and the late-stage hadronic interactions (simulated by UrqMD) and thus is an excellent probe for the initial state of the heavy ion collisions. Here, we have shown that $\NSC(2,3)$ can be used to discriminate the different nuclear configurations in ${}^{16}$O+${}^{16}$O collisions, particularly in the ultra-central heavy ion collisions~\cite{Li:2025hae}. As shown in Fig.~\ref{plot:NSC_23}, $\NSC(2,3)$ in the initial state (i.e, correlations between $\varepsilon_2$ and $\varepsilon_3$, shown by dashed lines), very well discriminates the Woods--Saxon (black dashed) and the three different tetrahedral $\alpha$ clustered nuclear configurations. In the ultra-central collisions, for example, 0-0.1\% centrality class, Cl. I (dashed red), Cl.II (dashed green), and Cl. III (dashed blue) values for $\NSC(2,3)$ are enhanced by more than 100\%, 300\%, and 400\% compared to that of Woods--Saxon, respectively. Similarly, $\NSC(2,3)$ in the final state (i.e, the correlations between $v_2$ and $v_3$, shown by solid lines) is able to distinguish different initial states. In 0-1\% centrality class, $\NSC(2,3)$ in the final state of Cl. I (red solid), Cl. II (green solid), and Cl. III (blue solid) are enhanced by about 37\%, 98\%, and 123\% compared to Woods--Saxon, respectively. However, this clear distinction is lost as one approaches mid-central and peripheral collisions. This may be because, in small systems like ${}^{16}$O+${}^{16}$O there is insufficient flow in the mid-central and peripheral collisions as can be seen from suppression of $\langle\langle p_T \rangle\rangle$ at these centralities in Fig.~\ref{plot:mean_pt}. In heavy ion collisions, $\langle\langle p_T \rangle\rangle$ remains almost flat throughout all centrality classes~\cite{STAR:2008med,ALICE:2018hza}. The bands around the final state results (solid lines)  show systematic uncertainties due to shear and bulk viscosities. The major contribution to the systematics band is due to bulk viscosity and the influence of shear viscosity is negligible.

\par
\begin{figure}[t!]
\includegraphics[width=0.45\textwidth]{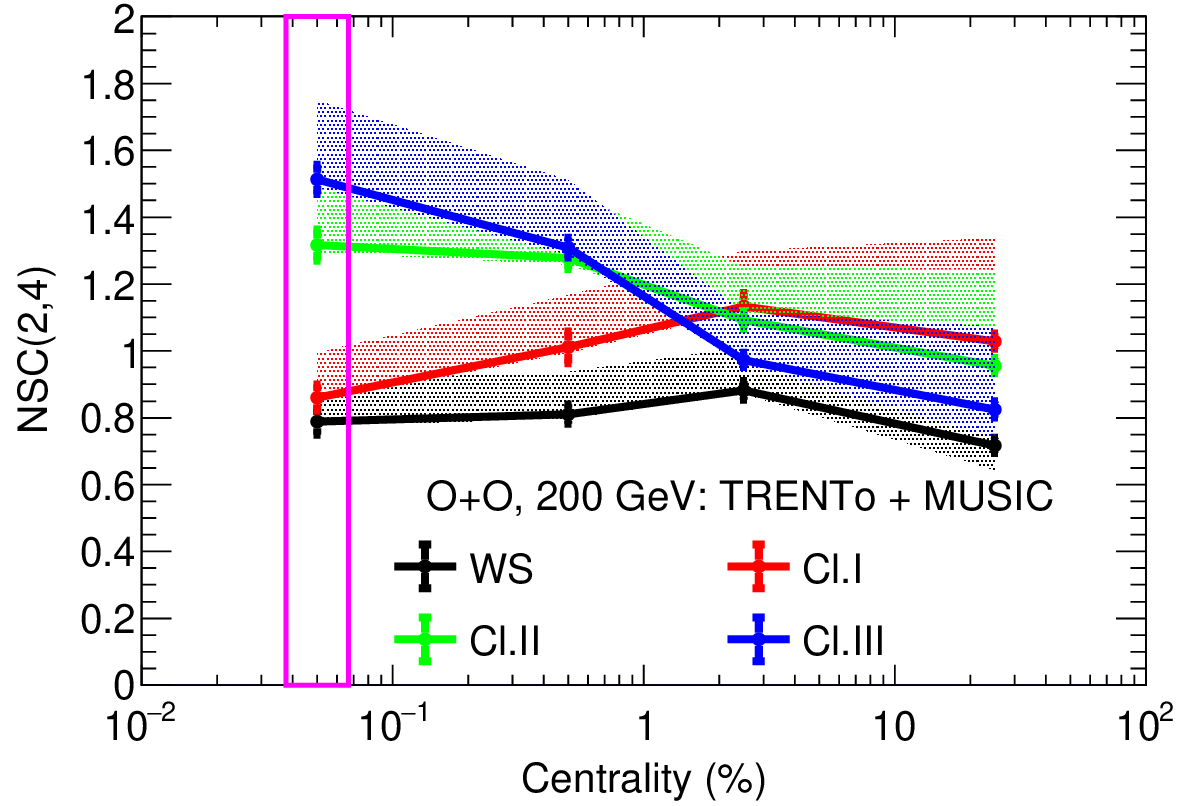}
\caption{\label{plot:NSC_24} Normalized symmetric cumulant vs centrality in ${}^{16}$O+${}^{16}$O collisions at $\sqrt{s_{NN}}$ = 200 GeV for the TRENTo+MUSIC framework. The shaded band represents the systematic uncertainties associated with $\eta/s$ and $\zeta/s$, with the dominant contribution arising from $\zeta/s$ and the influence of $\eta/s$ being comparatively smaller. $\NSC(2, 4)$ is consistently positive.}
\end{figure}

Similarly, \(\NSC(2,4)\), as shown in Fig.~\ref{plot:NSC_24}, can be used to distinguish between different nuclear configurations; however, it exhibits larger systematic uncertainties due to its sensitivity to shear and bulk viscosities, as well as to hadronic rescattering effects modeled by UrQMD~\cite{Shafi:2024tbr}. The additional evolution through UrQMD may significantly modify the results, potentially diminishing the discriminatory power of this observable with respect to different nuclear configurations.

\par
\begin{figure}[t!]
\includegraphics[width=0.45\textwidth]{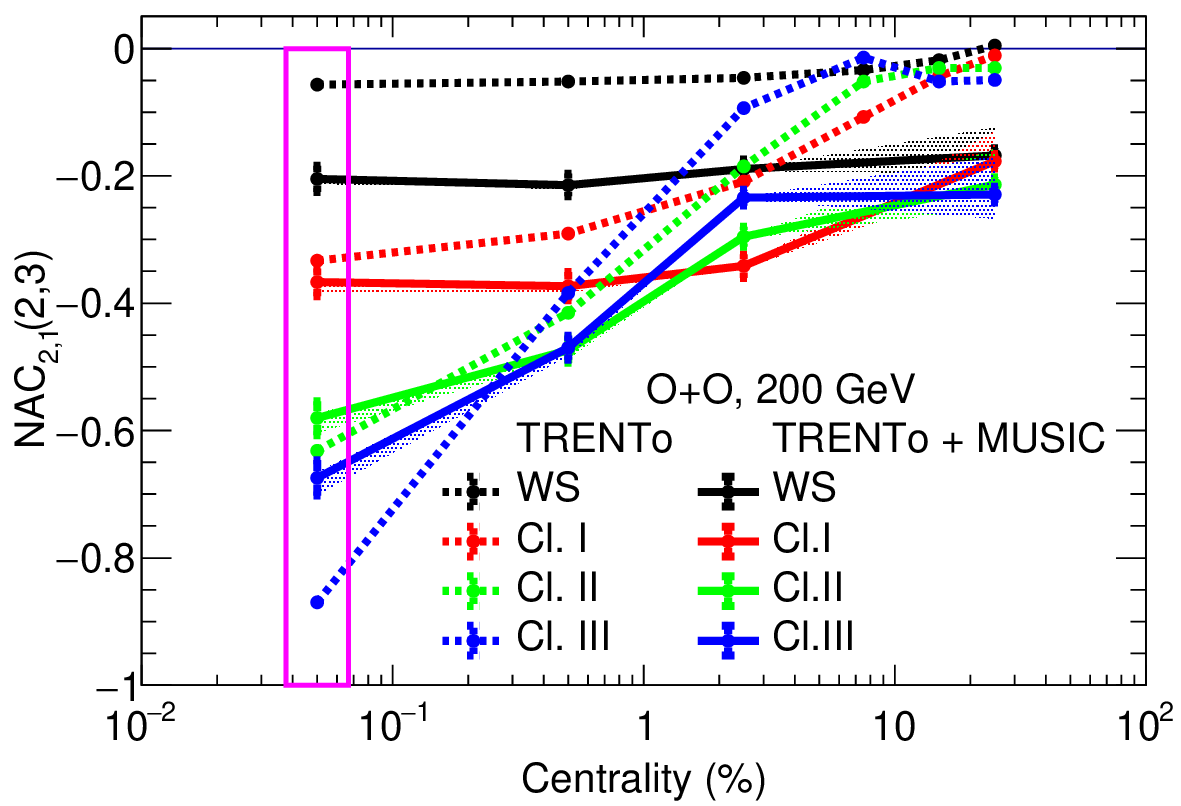}
\caption{\label{plot:NAC_21_23} Normalized asymmetric cumulant vs centrality in ${}^{16}$O+${}^{16}$O collisions at $\sqrt{s_{NN}}$ = 200 GeV. The shaded bands represent systematic uncertainties associated with $\eta/s$ and $\zeta/s$, dominated by the latter in non-central collisions. $\NAC_{2,1}(2, 3)$ is consistently negative. Dashed lines represent results from TRENTo, while solid lines correspond to the TRENTo+MUSIC framework.}
\end{figure}

We now present the results for the asymmetric cumulants \( \NAC_{2,1}(m,n) \). In Fig.~\ref{plot:NAC_21_23}, we also illustrate the distinction among different nuclear configurations based on \( \NAC_{2,1}(2,3) \). We found that for \( \NAC_{2,1}(2,3) \), in the ultra-central collisions, for example, 0-0.1\% centrality class, Cl. I (red dashed), Cl.II (green dashed), and Cl. III (blue dashed) values are enhanced by around 80\%, 180\%, and 230\% compared to Woods--Saxon, respectively. This distinction arises from the initial state as can be seen from the large differences in the initial states values of \( \NAC_{2,1}(2,3) \). Again, this clear distinction is washed out as one approaches mid-central and peripheral collisions. The bands show the systematics due to shear and bulk viscosities. For \( \NAC_{2,1}(2,3) \), the narrow bands imply minuscule effect of both bulk and shear viscosity on this observable. This makes it even more effective than 
$\NSC(2,3)$ to constraint the nuclear geometry of small nuclei like ${}^{16}$O.

\par
\begin{figure}[t!]
\includegraphics[width=0.45\textwidth]{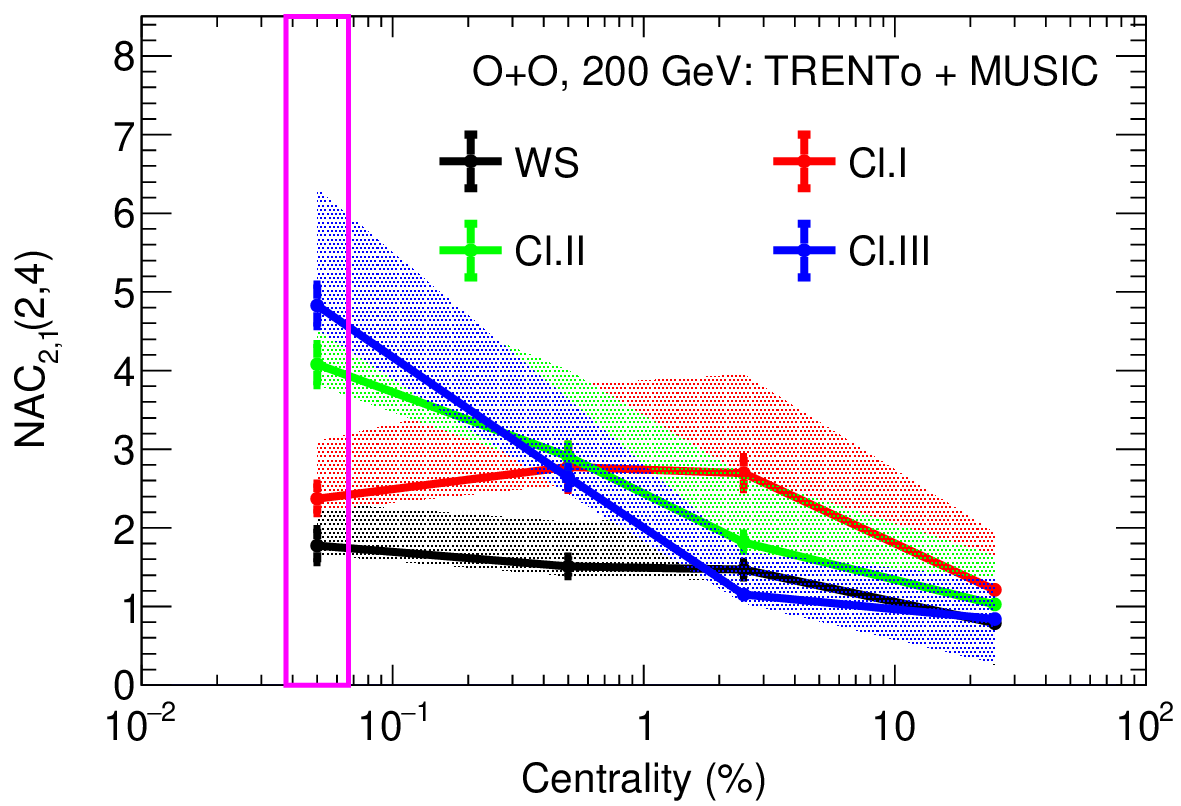}
\caption{\label{plot:NAC_21_24} Normalized asymmetric cumulant vs centrality in ${}^{16}$O+${}^{16}$O collisions at $\sqrt{s_{NN}}$ = 200 GeV for the TRENTo+MUSIC framework. The shaded band represents the systematic uncertainties associated with $\eta/s$ and $\zeta/s$, with the dominant contribution arising from $\zeta/s$ and the influence of $\eta/s$ being comparatively smaller. $\NAC_{2,1}(2, 4)$ is consistently positive.}
\end{figure}

The influence of transport coefficients—specifically shear and bulk viscosities—as well as resonance decays and late-stage hadronic interactions, is particularly pronounced for \(\NAC_{2,1}(2,4)\)~\cite{Shafi:2024tbr}. This is reflected in Fig.~\ref{plot:NAC_21_24}, where the uncertainty bands are noticeably wider. While we are still able to distinguish between the Woods–Saxon and Cl. III configurations in the most ultra-central collisions, it is important to note that the inclusion of hadronic rescattering via UrQMD, which has not been applied in this analysis, may significantly alter the results and reduce the discriminatory power of \(\NAC_{2,1}(2,4)\) with respect to different nuclear configurations. 

\par
In Ref.~\cite{Shafi:2024tbr}, we showed the noteworthy similarities in the magnitude and centrality dependence of the normalized cumulants at RHIC and LHC. The comparison indicates minimal energy dependence for the normalized cumulants. Therefore, our results are also applicable at LHC energies also.

\par
As a result, the normalized symmetric and asymmetric cumulant measurements possess the capability to distinguish between diverse nuclear geometries, particularly within hydrodynamic and transport frameworks. \(\NSC(2,3)\) and \(\NAC_{2,1}(2,3)\), due to their insensitivity to hydrodynamic model parameters, resonance decay, and late-stage hadronic interactions, are especially valuable for constraining the initial conditions of the system's evolution.

\subsection{$v_1^{even}$}
\begin{figure}[t]
\includegraphics[width=0.45\textwidth]{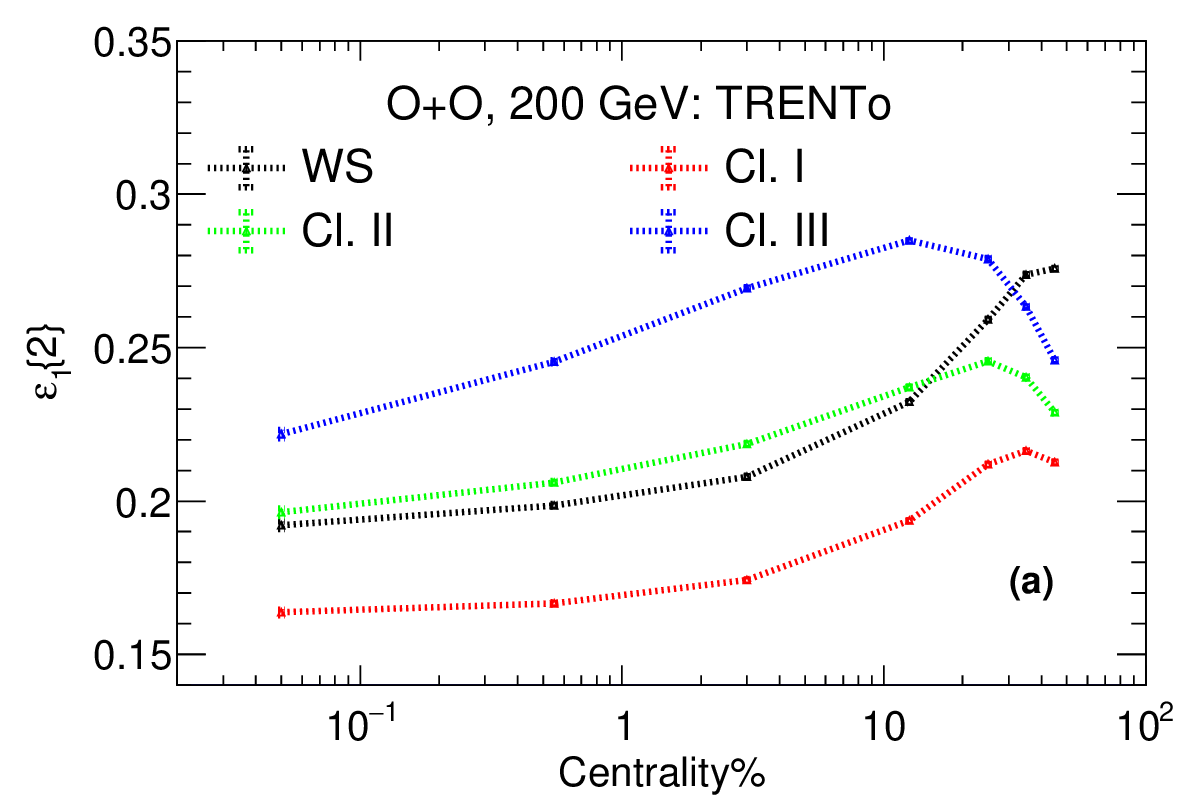}\\
\includegraphics[width=0.45\textwidth]{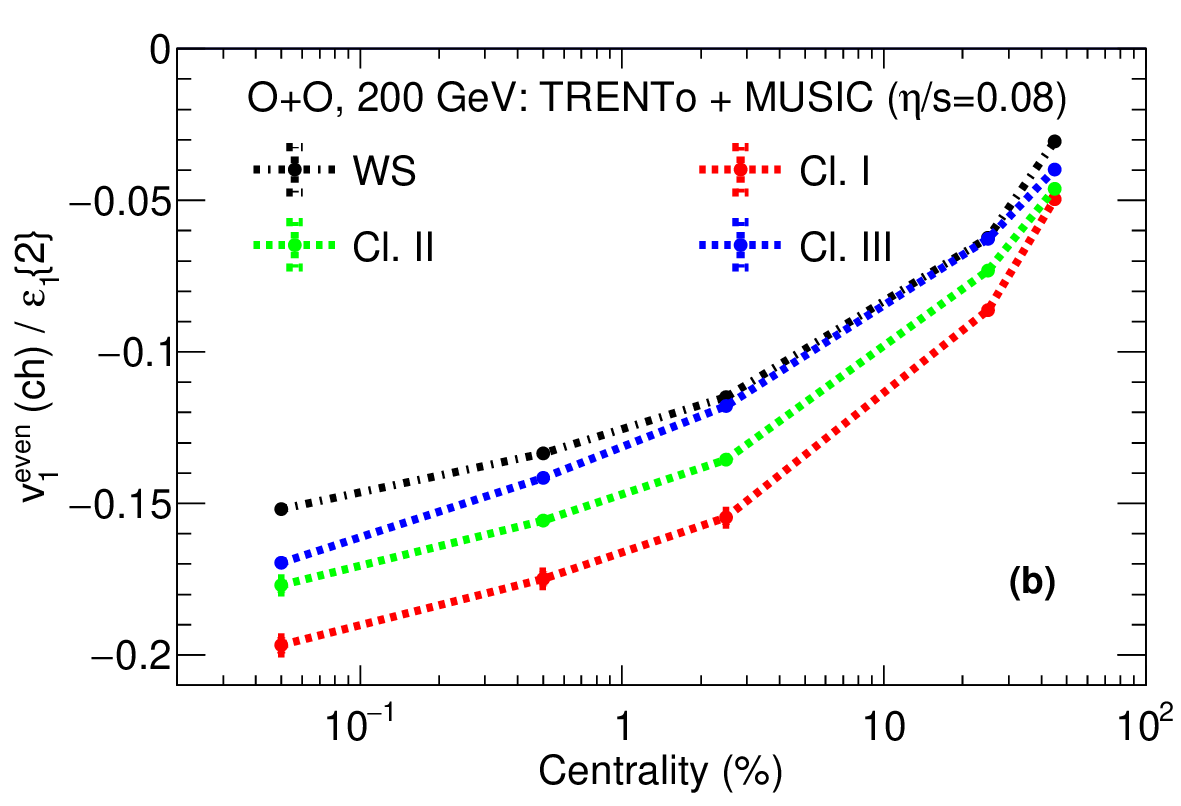}\\
\includegraphics[width=0.45\textwidth]{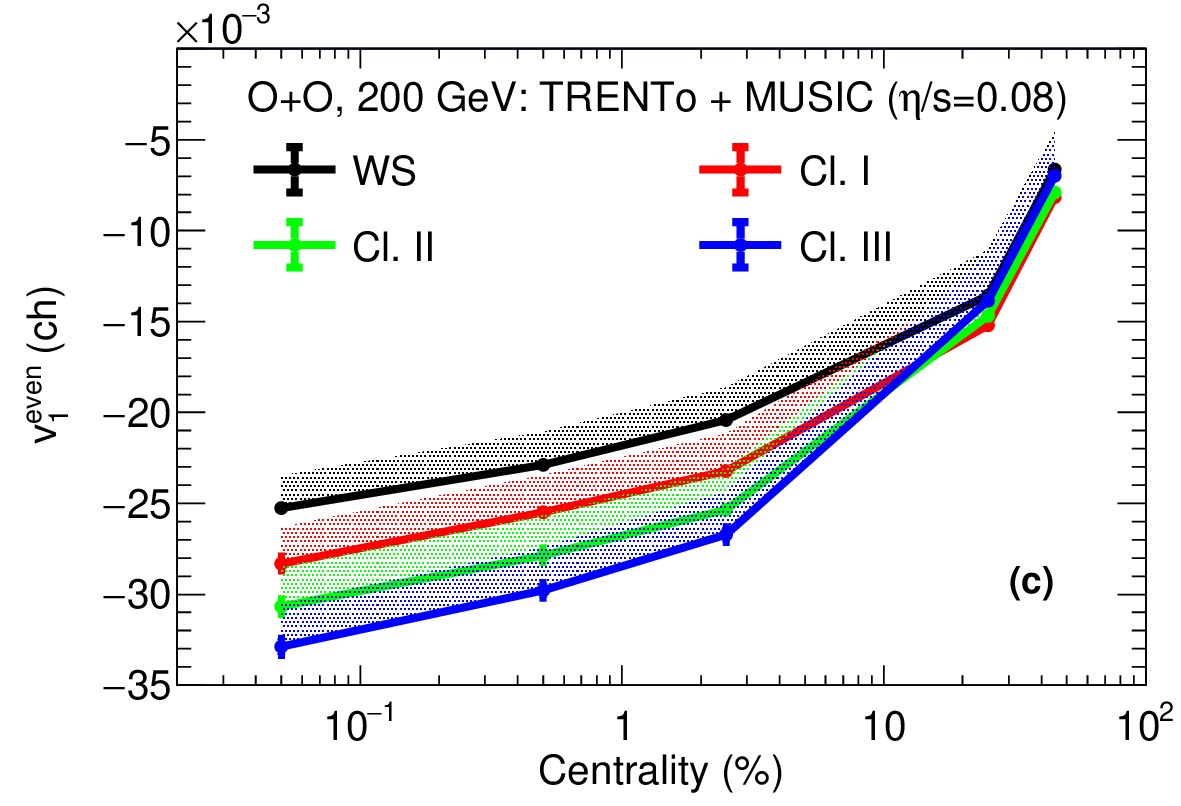}
\caption{\label{plot:v1even} Panel (a) shows $\varepsilon_1\{2\}$, panel (b) shows $v_1^{even}/\varepsilon_1\{2\}$ and panel (c) shows $v_1^{even}$ in ${}^{16}$O+${}^{16}$O collisions at $\sqrt{s_{NN}}$ = 200 GeV for the TRENTo+MUSIC framework. The shaded band for $v_1^{even}$ represents the systematic uncertainties associated with $\eta/s$.}
\end{figure}

In Fig.~\ref{plot:v1even}(a), we show how \(\varepsilon_1\{2\}\) varies with centrality for different nuclear configurations. A clear hierarchy is observed among the clustered configurations, with the configuration featuring more compact clusters exhibiting larger \(\varepsilon_1\{2\}\). This behavior arises because, as the clusters become more compact, the number of effective sources decreases, leading to larger fluctuations. To evaluate the sensitivity of \(v_1^{\text{even}}\) to different nuclear configurations, we compare its values for the configurations listed in Table~\ref{tab:ic_par}, using the first two parameter sets described in Table~\ref{table:parsets}, as shown in Fig.~\ref{plot:v1even}(c). Although \(\varepsilon_1\{2\}\) increases from central to peripheral collisions, both the response ratio \(v_1^{\text{even}}/\varepsilon_1\{2\}\), as shown in Fig.~\ref{plot:v1even}(b), and the mean transverse momentum \(\langle\langle p_T \rangle\rangle\), as shown in Fig.~\ref{plot:mean_pt}, decrease correspondingly, which explains the observed centrality dependence of \(v_1^{\text{even}}\). We assessed the sensitivity of \(v_1^{\text{even}}\) to hadronic rescattering using UrQMD and observed negligible impact. Regarding the opposite hierarchy of \(v_1^{\text{even}}\) and the linear response across different clusters, two primary factors drive \(v_1^{\text{even}}\): the initial dipole asymmetry, \(\varepsilon_1\), and the hydrodynamic response of the QGP medium. Although \(\varepsilon_1\) increases from Cl.~I to Cl.~III, the response (as well as \(\langle\langle p_T \rangle\rangle\)) follows a decreasing trend. The dominance of \(\varepsilon_1\) results in a similar hierarchy for \(v_1^{\text{even}}\) across the clusters. As the clustering becomes more compact, the magnitude of \(v_1^{\text{even}}\) increases. We observe substantial variation in \(v_1^{\text{even}}\) across different tetrahedral \(\alpha\)-clustered O+O configurations, with a change of approximately 30\% between the Woods--Saxon and Cl.~III cases. This makes \(v_1^{\text{even}}\) a promising observable for distinguishing between Woods--Saxon and \(\alpha\)-clustered nuclear geometries.

\section{Conclusions}
We have presented results for multi-particle correlation functions in heavy-ion collisions at top RHIC energy using a hybrid framework based on TRENTo model, and MUSIC viscous hydrodynamics simulations.
\par
First, we adjusted the free parameters, such as shear and bulk viscosities, to describe particle multiplicities, mean transverse momentum, and anisotropic flow for Au+Au collision system. After that, we discussed the impact of different nuclear geometries on the novel observables considering their sensitivities to shear and bulk viscosities, resonance decay, and hadronic interactions.
\par
We have studied correlators that measure the correlations between flow harmonics of varying orders, including both symmetric and asymmetric cumulants as functions of centrality at midrapidity in ${}^{16}$O+${}^{16}$O collisions at $\sqrt{s_{NN}}=200$ GeV. These observables are very sensitive to initial nuclear geometry. We found that \(\NSC(2, 3)\) and in particular \(\NAC_{2,1}(2, 3)\) being insensitive to both the hydro model parameters (shear and bulk viscosity) and late-stage hadronic interactions, are good observables for distinguishing different nuclear distributions in light nuclei like ${}^{16}$O. This strongly supports their effectiveness as reliable tools for constraining the initial state of the system’s evolution. Conversely, \(\NSC(2,4)\) and \(\NAC_{2,1}(2,4)\) exhibit considerable sensitivity to the hydrodynamic model parameters and to hadronic rescattering effects such as those modeled by UrQMD. Although in the absence of late-stage hadronic interactions these observables are capable of distinguishing between the Woods--Saxon and the most tetrahedral configurations in ultra-central collisions, the inclusion of such effects may significantly alter the results, potentially reducing their discriminatory power with respect to different nuclear configurations.
\par
We observed an anticorrelation between the event-by-event fluctuations of \( v_2 \) and \( v_3 \) up to central collisions, while the fluctuations of \( v_2 \) and \( v_4 \) are found to be positively correlated. The observed anti-correlation between \( v_2 \) and \( v_3 \) appears to be driven by the initial-state anti-correlation between \( \varepsilon_2 \) and \( \varepsilon_3 \), supporting the notion of linearity between \( \varepsilon_n \) and \( v_n \)~\cite{Alver:2010gr}. This is consistent with the hydrodynamic picture in which final-state fluctuations originate from initial-state geometry. In contrast, the correlation between \( v_2 \) and \( v_4 \) is primarily attributed to the nonlinear hydrodynamic response of the medium, as the linear component of the \( v_4 \) response is comparatively weaker.
\par
In heavy-ion collisions, \(\NSC(m,n)\) measured by the STAR collaboration for Au+Au collisions at \(\sqrt{s_{NN}} = 200\) GeV and the ALICE collaboration for Pb+Pb collisions at \(\sqrt{s_{NN}} = 5.02\) TeV exhibit similar magnitudes and centrality trends. For \(\NAC_{2,1}(m,n)\), previous model results for Au+Au at \(\sqrt{s_{NN}} = 200\) GeV showed behavior similar to the measurements made by ALICE. We anticipate comparable behavior in light-ion collisions, such as \({}^{16}\text{O}+{}^{16}\text{O}\) collisions.
\par
We also demonstrate that \( v_1^{\text{even}} \) proves to be a sensitive observable for distinguishing between Woods--Saxon and \( \alpha \)-clustered nuclear geometries. The centrality dependence of \( v_1^{\text{even}} \), explained by \( \varepsilon_1 \) and changes in the response ratio, \( v_1^{\text{even}}/\varepsilon_1\{2\} \), and mean transverse momentum, \( \langle\langle p_T \rangle\rangle \), reveals a clear trend from central to peripheral collisions. Moreover, substantial variation in \( v_1^{\text{even}} \) is observed across different tetrahedral \(\alpha\)-clustered O+O configurations, with a notable 30\% difference between Woods--Saxon and \( \alpha \)-clustered distributions, with more compact clustering leading to an increased deviation from the Woods–Saxon nuclear geometry baseline. These findings highlight the potential of \( v_1^{\text{even}} \) as a tool for probing nuclear structure and geometry. Furthermore, we found that hadronic rescattering, as modeled by UrQMD, has a negligible impact on the final-state \( v_1^{\text{even}} \) signal.

\section{Acknowledgements}
We gratefully acknowledge Tribhuban Parida for the valuable discussions and for providing the computational setup, which played a crucial role in facilitating our analysis by enabling us to carry out extensive simulations and test various scenarios efficiently.

\section{Data Availability}
The data and code that support the findings of this article are not publicly available. The data and code are available from the authors upon reasonable request.

\bibliography{main}

\end{document}